\documentclass[twocolumn,amsmath,amssymb,amsfonts,aps,prl,showpacs,preprintnumbers,superscriptaddress]{revtex4-1}
\usepackage{graphicx}
\usepackage{dcolumn}
\usepackage{bm}
\usepackage{color}

\begin{document}

\title{Ultrafast Dynamics of the Charge Density Wave State in Layered Cu$_x$TiSe$_2$}

\author{D.B. Lioi}
\affiliation{Department of Physics, Drexel University, Philadelphia, Pennsylvania 19104, USA}

\author{R.D. Schaller}
\affiliation{Center for Nanoscale Materials, Argonne National Laboratory, Argonne, Illinois 60439, USA}

\author{G.P. Wiederrecht}
\affiliation{Center for Nanoscale Materials, Argonne National Laboratory, Argonne, Illinois 60439, USA}

\author{G. Karapetrov}
\affiliation{Department of Physics, Drexel University, Philadelphia, Pennsylvania 19104, USA}

\date{\today}
\tighten
\begin{abstract}
We report on a transient optical reflectivity study of the charge density wave (CDW) in Cu$_x$TiSe$_2$ single crystals. Our measurements reveal that the system undergoes a quantum phase transition at x=0.04 from a strong commensurate CDW (x$\le$0.04) to a soft incommensurate CDW (x$\ge$0.04). We find that the cooperative driving mechanisms for the commensurate CDW, the excitonic insulator mechanism and the soft L$_1$$^-$ phonon mode, decouple at x=0.04 with the observed fluctuations in the folded Se-4p band dominating the transition. We also demonstrate a loss of coherence in the A$_{1g}$ phonon with increased copper intercalation of the parent lattice, indicating a loss of long-range lattice order. These findings provide compelling evidence that TiSe$_2$ undergoes a quantum phase transition upon copper intercalation from a state of commensurate charge order to a state with a different symmetry in which the new charge order coexists with the superconducting phase.

\end{abstract}

\pacs{}
\keywords{charge density wave, TiSe$_2$, quantum phase transition, superconductivity}

\maketitle
Development of different phases in a material can be described by the emergence of a corresponding order parameter that is triggered by a sudden change in symmetry, as first noted by L.\,D.\,Landau.~\cite{Landau_PhaseTrans_1937}  At temperatures close to 0\,K, crossing a boundary between two phases means that something fundamentally has changed in the quantum mechanical ground state of the system, i.e. the system has undergone a quantum phase transition (QPT). In this case the change in the state is not due to the temperature, but due to change in some parameter in the Hamiltonian describing the system, for example strain, doping, disorder, etc. The quantum fluctuations associated with QPTs are also driven by the quantum mechanics on a microscopic scale, unlike the long range critical fluctuation in classical (thermally driven) phase transitions. The importance of QPTs goes far beyond the zero-temperature case: QPT determines the coexistence of different phases in the material at finite temperatures and, in fact, provides answers about the coexistence and competition of different order parameters over a large area of the phase diagram. Thus, studies of QPTs in materials with coexisting order parameters could expose the mechanisms of correlated electron states at finite temperatures such as charge density wave (CDW), high-temperature superconductivity, etc.

Many high-temperature superconductors have a strong intrinsic disorder making it difficult to precisely control the physical parameter that triggers the QPT. Transition metal dichalcogenides (TMDs), on the other hand, provide an excellent model system with relatively simple crystalline structure and a wealth of ground states that result in the coexistence of charge order, Cooper pairs, excitonic condensate, etc. Among the many TMDs exhibiting excitonic properties~\cite{Liang_1972_JoPC,Lin_2014_NanoLet,Ross_2013_NatureCom,Huang_2016_ScientificRep}, TiSe$_2$ is a prime candidate for displaying the excitonic insulator behavior~\cite{Kohn_1967_PRL} in which a CDW phase arises from strong Coulomb interactions. The excitonic condensate was proposed as the driving mechanism for the commensurate CDW in TiSe$_2$~\cite{Wilson_1977_SSC,Wilson_1978_PSSB,Salvo_1976_PRB} in part due to its low carrier density, which leads to reduced screening between the holes in the Se-4p band and electrons in the Ti-3d band. However, it has been difficult to confirm that excitonic insulator mechanism is the dominant reason for triggering the CDW in TiSe$_2$.~\cite{Kogar_EELS_arxiv2016}

The primary competing hypothesis to the purely excitonic origin of the CDW in TiSe$_2$ has been the band-type Jahn-Teller mechanism combined with electron-phonon coupling~\cite{Hughes_1977_JoPC,Whangbo_1992_JACS,Motizuki_1981_PhysicaBC,Motizuki_1981_SSC,Rossnagel_2002_PRL}. The discovery of a soft phonon at the T$_{CDW}$=200\,K~\cite{Holt_2001_PRL,Weber_2011_PRL} lent weight to the Jahn-Teller picture~\cite{Motizuki_1981_PhysicaBC,Motizuki_1981_SSC}. With both the excitonic insulator~\cite{Salvo_1976_PRB,Cercellier_2007_PRL,Pillo_2000_PRB,Hellmann_2012_NatureCom} and Jahn-Teller~\cite{Anderson_1985_SSC,Anderson_1985_SSC,Rossnagel_2002_PRL} views supported by various experiments, the question turned from whether the CDW transition is caused by either excitons \textit{or} by phonons interacting with electrons at the Fermi surface, to the more subtle question of how these two mechanisms \textit{cooperate}~\cite{Wezel_2010_PRB,Phan_2013_prb,Porer_2014_NatureMat,Zenker_2014_PRB} to give rise to the observed CDW in TiSe$_2$, including the recently observed chiral CDW phase~\cite{Ishioka_2010_PRL,Castellan_2013_PRL,Zenker_2013_PRB}. Recent ultrafast pump-probe~\cite{Vorobeva_2011_PRL,Ishioka_2010_PRL,Porer_2014_NatureMat} and transient angle resolved photo-emission spectroscopy~\cite{Rohwer_2011_Nature,Hellmann_2012_NatureCom,Rohde_2014_JESRP,Monney_2016_prb} experiments have been aimed at disentangling electron-electron and electron-phonon interactions in pristine TiSe$_2$. Up to now the exact relation between these mechanisms has remained elusive~\cite{Zenker_2014_PRB,Phan_2013_prb,Hellmann_2012_NatureCom,Porer_2014_NatureMat,Kogar_EELS_arxiv2016}.

\begin{figure*}[t]
	\includegraphics[width=16cm]{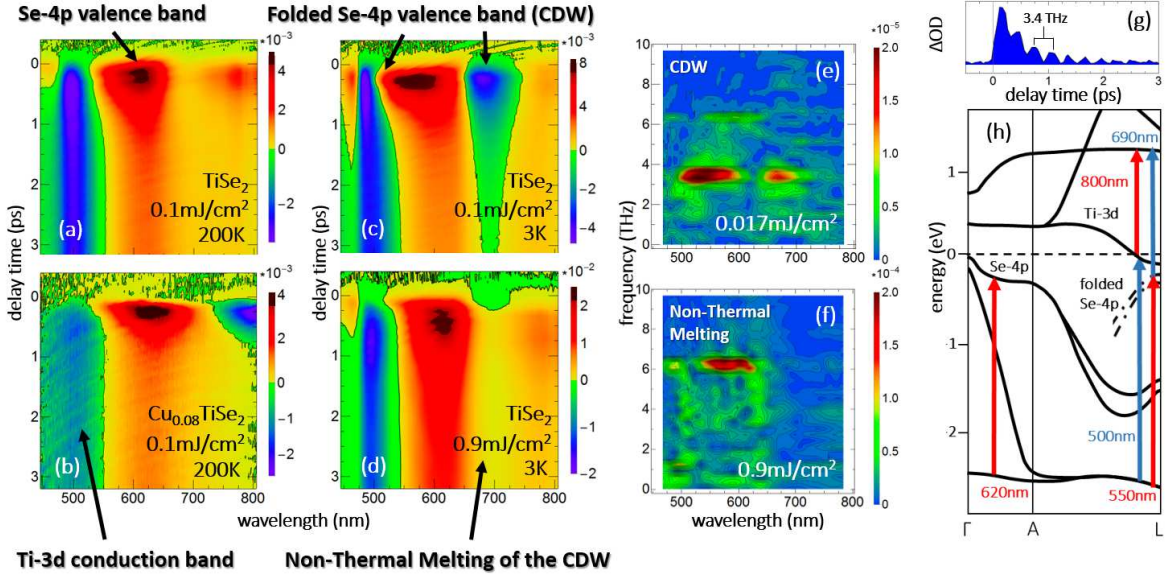}
	\caption{(a-d) Transient reflectivity change in optical density $\Delta$OD for (a) normal/semimetalic phase, (b) metallic phase, (c) CDW phase, (d) non-thermally melted CDW state (threshold at 0.25\,mJ/cm$^2$). Warm colors denote ESA while cool colors denote GSB. (e,f) Fourier transforms of coherent A$_{1g}$ oscillations for the CDW and normal phases. (g) time domain trace of TiSe$_2$ optical response at 3K and a 0.017\,mJ/cm$^2$ pump fluence. (h) TiSe$_2$ electronic band diagram adapted from Zunger et al~\cite{Zunger_1978_PRB} with arrows specifying optically excited resonances. Blue arrows mark GSB resonances and red arrows mark ESA resonances.  }
	\label{fig1}
\end{figure*}


The recent discovery of superconductivity in Cu-doped TiSe$_2$~\cite{Morosan_2006_NaturePhys} and its coexistence with the CDW~\cite{Iavarone_2012_PRB} has reinvigorated research on this system. Of particular interest has been the different nature of the CDW in TiSe$_2$ and Cu$_x$TiSe$_2$~\cite{Kogar_arxiv2016}, as well as in TiSe$_2$ under pressure.~\cite{Joe_2014_NaturePhys}
In this work we systematically tune the level of Cu intercalation in TiSe$_2$ and simultaneously track the changes in the excitonic and phonon subsystems with the goal of understanding the evolution of underlying ground states as the system transitions from a narrow-band semiconducting state (TiSe$_2$) to superconducting state in Cu$_x$TiSe$_2$ (at x\,$>$\,0.04).

We have performed ultrafast optical pump-probe experiments in which we have varied laser fluence, Cu doping, and temperature to explore the phase diagram of Cu$_x$TiSe$_2$ single crystals. To our knowledge this is the first such systematic study that dynamically monitors the electron-electron and electron-phonon interactions in Cu$_x$TiSe$_2$ as the system progresses from the normal and CDW phases into the metallic and superconducting phases. Copper doping was found to weaken the excitonic condensate and decouple it from the L$_1$$^-$ phonon mode, leading to fluctuations that are indicative of a quantum phase transition at x=0.04. This, along with a loss of coherence in the A$_{1g}$ response at Cu$_{0.04}$TiSe$_2$, supports the picture of a commensurate CDW in intrinsic TiSe$_2$ which transitions upon Cu intercalation into a state with a different symmetry characterized by an incommensurate CDW that coexists with superconductivity.

High quality single crystals were grown using the chemical vapor transport method~\cite{Salvo_1976_PRB,Oglesby_1994_JoCG,Wilson_1977_SSC} and characterized using energy dispersive X-ray spectroscopy, X-ray diffractometry, and variable temperature transport. The crystals were exfoliated in inert atmosphere to expose a clean surface before being mounted in the optical setup in reflection geometry. The pump pulse had a wavelength 1160\,nm (1.07eV), a 1\,kHz repetition rate, and 600\,$\mu$m diameter spot size. Excited carriers were dynamically monitored by a white light probe pulse. If the pump pulse removes a resonance within the electronic band structure it results in a negative spectral signal at that wavelength, i.e. ground state bleaching (GSB) while adding a resonance yields a positive signal, i.e. excited state absorption (ESA).

Fig.1(a) shows that the normal phase in pristine TiSe$_2$ has its most prominent change in optical density ($\Delta$OD) as a GSB signal at 500\,nm which is assigned to the Ti-3d conduction band, and an ESA signal at 620\,nm which is assigned to the Se-4p valence band along $\Gamma$-A. The magnitude of these signals increases with the pump fluence. Optically excited resonances within the electronic band structure change as the material enters the CDW phase at lower temperatures, as shown in Fig.1(c) with two prominent features appearing at 550\,nm and 690\,nm. These are assigned to the backfolded Se-4p valence band that forms as the sample enters the CDW phase~\cite{Rohwer_2011_Nature}.

To further demonstrate that the new 550\,nm and 690\,nm resonances in Fig.1(c) are consistent with the folded Se-4p band, the pump fluence was raised above the non-thermal melting threshold (0.25\,mJ/cm$^2$) of the CDW phase~\cite{Hellmann_2012_NatureCom,Rohwer_2011_Nature,Vorobeva_2011_PRL}. Fig.1(d) shows the resulting transient spectra within the first 100\,fs exhibiting a similar response as the low fluence CDW measurement in Fig.1(c). This is rapidly followed by the picosecond-nanosecond response of the transient spectra which appears to mimic the normal phase response shown in Fig.1(a). The 550\,nm and 690\,nm assignments are further corroborated by coherent oscillations of A$_{1g}$ phonons associated with the normal phase lattice (6.2\,THz) and the CDW lattice (3.4\,THz)~\cite{Vorobeva_2011_PRL,Snow_2003_PRL,Monney_2016_prb} as shown in Fig.1(g). FFT of these oscillations from the transient spectra reveals that the A$_{1g}$ CDW oscillation is centered at 550\,nm and 690\,nm, while the normal phase A$_{1g}$ resides outside of these regions and instead in the Se-4p band along $\Gamma$-A (Fig.1(e,f)). The pump pulse at high fluences ($\geq$ 0.9\,mJ/cm$^2$) is shown to cause non-thermal melting of the CDW which yields the normal phase A$_{1g}$ response shown  in Fig.1(d). Fig.1(h) illustrates the optically excited transitions at different energies within the electronic band structure of TiSe$_2$ in the CDW phase.

Having assigned the change in optical response of Cu$_x$TiSe$_2$ to optically excited resonances near the Fermi surface, we now turn to a systematic quantitative analysis of the transient spectra. This was accomplished with the open-source software package Glotaran~\cite{Snellenburg_2012_JoSS} which models the transient optical response as follows:

\begin{eqnarray}
Transient~Optical~Response =\nonumber\\
\sum_{i}
IRF
\otimes
\Delta OD_i(\lambda)
\times
e^{-t/\tau_i}\nonumber
\end{eqnarray}

In our case the instrument response function (IRF) was convoluted with two exponential decay functions, each with a starting magnitude ($\Delta$OD coefficient that is wavelength dependent) and decay constants ($\tau_1$ and $\tau_2$). The $\Delta$OD coefficient signifies an excited carrier population at the beginning of the exponential decay for each time regime. The process associated with $\tau_1$ begins when the excited carrier population has already self-scattered, and begins interacting with the lattice to further lower its energy toward the Fermi surface. The process characterized with decay rate $\tau_2$ is the one in which the electrons and holes situated as close to the Fermi surface as they can recombine across the bandgap on nanosecond timescales.

\begin{figure}[t]
	\includegraphics[width=6cm]{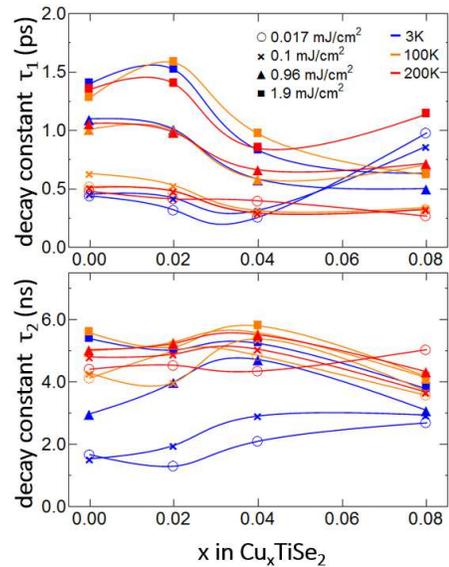}
	\caption{Fitted decay constants across all parameters. $\tau_1$ (ps) is dominated by electron-phonon interactions with contributions from changing electronic band structure. $\tau_2$\,(ns) is governed by electron-hole recombination across the bandgap. Blue symbols in the right-most column lie within the superconducting part of the phase diagram. Curves are shown to aid the eye.}
	\label{fig2}
\end{figure}

The shorter time scale decay constant $\tau_1$ is governed by electron-phonon coupling. Typically longitudinal optical (LO) phonons are the dominant contributor to the energy reduction of the excited electronic carriers. Stronger electron-phonon coupling leads to faster decay rates (smaller $\tau_1$ values) assuming static electronic band structures and moderate excitation temperatures. However, the experiments on intrinsic TiSe$_2$ and low doped Cu$_x$TiSe$_2$ do not always satisfy these two conditions. A periodic lattice distortion (PLD)~\cite{Salvo_1976_PRB}, is a coherent alteration of the lattice from normal to superlattice state (with doubling of the size of the unit cell) that accompanies the formation of the CDW state. When large populations of carriers are excited during non-thermal melting at high fluences, the lattice reverts from the PLD state back to the normal phase~\cite{Hellmann_2012_NatureCom,Rohwer_2011_Nature,Vorobeva_2011_PRL}, temporarily changing the electronic band structure (i.e. the backfolded bands). The lattice, now in its excited normal phase, begins to revert back to the initial PLD state~\cite{Porer_2014_NatureMat} via emission of the soft  L$_1^-$ phonon~\cite{Holt_2001_PRL,Weber_2011_PRL} which is expected to dominate electron-phonon coupling~\cite{Motizuki_1981_PhysicaBC,Motizuki_1981_SSC}. This process of melting and subsequent refreezing of the  PLD phase slows the recovery time of the electronic carriers at picosecond timescales as they react to the changing energetic landscape for the duration of this process. Thus $\tau_1$, being a measure of electron-phonon coupling, in our case is strongly influenced by the lattice and structural phase transition at T$_{CDW}$.
\begin{figure*}[t]
	\includegraphics[width=14cm]{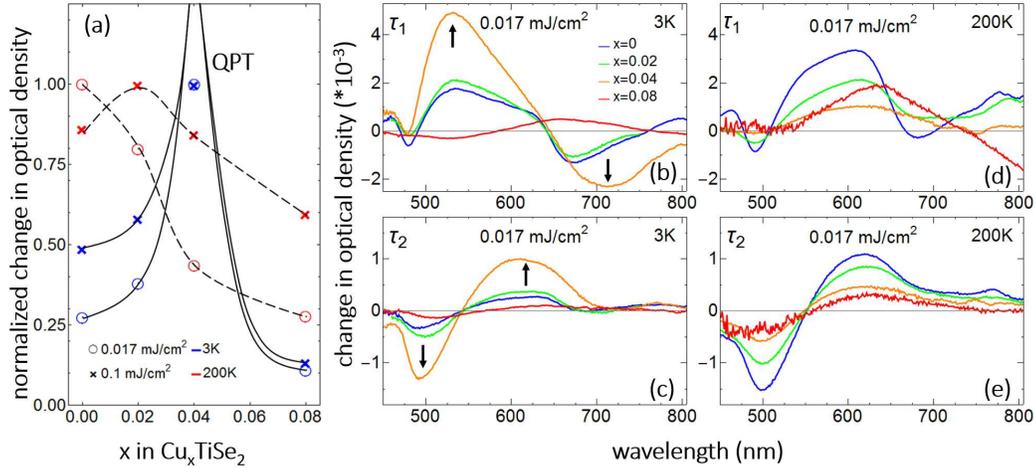}
	\caption{(a) $\Delta$OD coefficients associated with $\tau_2$ that were fit to the transient optical response of Cu$_x$TiSe$_2$ at 620\,nm. (b,d) global $\Delta$OD coefficients associated with picosecond timescales $\tau$$_1$ (c,e) and nanosecond timescales $\tau_2$ for the lowest beam fluence and varying temperatures. Color (online) shows the effect of increasing Cu doping. Arrows indicate fluctuations at the superconducting quantum phase transition. Color labels are the same throughout.}
	\label{fig3}
\end{figure*}

Fig.2 shows that the electron-phonon relaxation time $\tau_1$ is strongly dependent on the pump fluence, but relatively insensitive to the temperature for x$<$0.04. As a result, the lattice in pure TiSe$_2$ is slow to recover from large optical perturbations, consistent with the findings by Monney~\cite{Monney_2015_PRL} and Zenker~\cite{Zenker_2014_PRB} in regards to coupling between excitons and the soft phonon in TiSe$_2$. In their description the parent lattice becomes susceptible to excitonic condensation near T$_{CDW}$. As one lowers the temperature the excitonic condensate becomes robust below temperatures corresponding to the excitonic binding energy~\cite{Pillo_2000_PRB}. In principle, the condensate should form an excitonic insulator state if the lattice is static. In TiSe$_2$, instead, the softening of the lattice, which comes from the strong coupling of the condensing excitons to the L$_1^-$ mode, produces a PLD that prevents the excitonic condensate from forming. Above the Cu doping of x=0.04, $\tau_1$ becomes only weakly dependent on fluence. This signifies a decoupling of electronic carriers from the L$_1^-$ phonon and reduction of the PLD at higher Cu doping~\cite{Bussmann_2009_PRB}.

In order to explore the evolution of the excitonic side of CDW formation in Cu$_x$TiSe$_2$ with Cu concentration $x$, we turn to the decay constant $\tau_2$, which is determined by the electron-hole recombination rate. Deep within the CDW phase pristine TiSe$_2$ is unstable to the formation of excitons that results in a large spectral weight being transferred to the backfolded Se-4p band and strengthening the long range commensurate CDW order~\cite{Cercellier_2007_PRL}. The presence of the backfolded band opens a direct interband transition between the partially filled Ti-3d conduction band and the folded Se-4p band. This transition channel allows the excited carriers to quickly recombine, resulting in a small $\tau_2$ values at low pump fluences and low temperatures (3K), as shown in Fig.2. As we introduce Cu intercallant and approach the superconducting region of the phase diagram near x=0.04, the low temperature recombination rates increase, gradually converging to those observed at higher temperatures. This could be interpreted as a reduction of electronic density in the backfolded Se-4p band~\cite{Zhou_2007_PRL} which in turn indicates a weakening of the excitonic insulator mechanism and a loss in long range electronic order in the CDW superlattice~\cite{Cercellier_2007_PRL}.

The weakening of the excitonic CDW mechanism and the reduced fluence dependence of $\tau_1$ at x=0.04 and above are most likely related. If the soft L$_1^-$ mode couples with excitons, whose constituent holes and electrons lie at either end of the phonon vector, and these two mechanisms cooperate to form the CDW phase~\cite{Zenker_2014_PRB,Wezel_2010_PRB}, then weakening of the excitonic coupling between these electrons and holes will decrease the electron-phonon coupling via the L$_1^-$ mode. As a result the electronic bands will be less susceptible to renormalization in response to the pump pulse, removing the large dependence of $\tau_1$ on the pump fluence.

\begin{figure*}[t]
	\includegraphics[width=16.5cm]{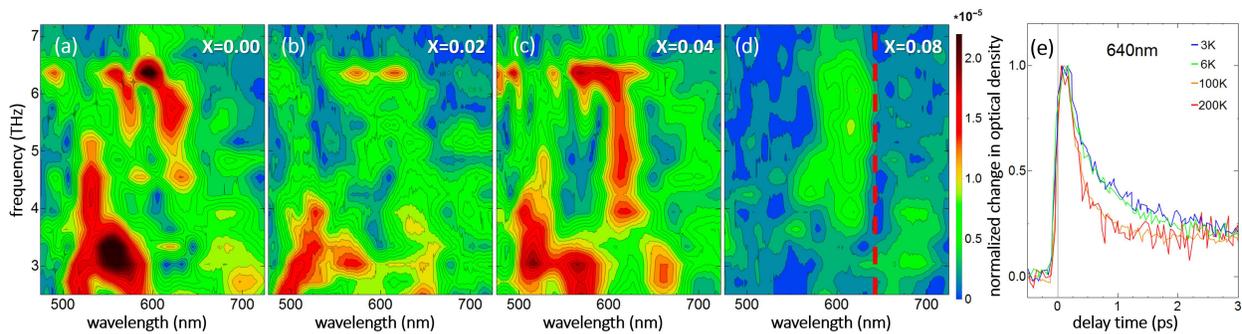}
	\caption{(a-d) FFTs of coherent phonon oscillations in the global transient spectra of Cu$_x$TiSe$_2$ for various Cu dopings at 3\,K and 0.1\,mJ/cm$^2$. (e) Select transient spectra for Cu$_{0.08}$TiSe$_2$ in the superconducting and metallic phases. The red line in part (d) denotes the optically excited probe resonance used in part (e).}
	\label{fig4}
\end{figure*}

When both of the excitonic and electron-phonon mechanisms discussed above have become sufficiently decoupled from each other by Cu intercalation, the superconducting phase begins to appear at T$\sim$0\,K near x=0.04. At that point in T$_c$-x phase diagram we observe clear signatures of the quantum phase transition as fluctuations in the transient optical response enhance dramatically. Fig.3(a) summarizes the experimental findings by showing non-monotonous behavior in the $\Delta$OD coefficient at 620\,nm with respect to Cu concentration $x$ near the transition from strong commensurate CDW into the superconducting phase with incommensurate CDW.~\cite{Kogar_arxiv2016} Fig.3(b-e) details the temperature evolution of the global $\Delta$OD coefficients associated with picosecond ($\tau_1$) and nanosecond ($\tau_2$) timescales at the lowest pump fluence used in these experiments. A large change in optical density $\Delta$OD signifies a large excited carrier population. Fig.3(b) shows that at 3K the $\Delta$OD coefficient associated with $\tau_1$ exhibits a nonmonotonous behavior, with a strong instability to small optical perturbations at the quantum phase transition at x=0.04. The transition at x=0.04 and T$\sim$0 separates the $\Delta$OD response of the robust commensurate CDW phase (at x$\le$0.04) from a vastly different $\Delta$OD response observed in the coexisting incommensurate CDW and superconducting state (at x$\ge$0.04).
The fingerprint of the instability at x=0.04 is the fluctuations in the population of the folded Se-4p band. These fluctuations are monitored by the 550\,nm and 690\,nm probe resonances (Fig.1(h)) at picosecond timescales (dominated by electron-phonon interactions, Fig. 3(b)), and by the 500\,nm and 620\,nm probe resonances at nanosecond timescales (recombination process, Fig.3(c)). The latter monitor the excited electrons at the bottom of the Ti-3d band and the holes at the top of the Se-4p band, respectively.

The above enhancement of fluctuations in $\Delta$OD at x=0.04 also have a strong signature in the coherent phonon response observed in Cu$_x$TiSe$_2$. FFTs of coherent electronic band oscillations are displayed in Fig.\,4, showing the evolution of the coherent A$_{1g}$ LO phonons for both the normal (6.2\,THz) and CDW (3.4\,THz) lattice modes with increasing Cu concentration. As mentioned above, the A$_{1g}$ phonons selectively couple to the electronic band structure, and are indicative of the long range lattice order~\cite{Coppersmith_1984_PRB,Mann_2016_prb}. In TiSe$_2$ and Cu$_{0.02}$TiSe$_2$ the FFT of these coherent electronic band oscillations corresponds to the 6.2\,THz A$_{1g}$ signal appearing as two lobes within the optical excited resonances that monitor the Se-4p band along the $\Gamma$-A direction, which we assign to the selenium 4p$_{x,y}$ and 4p$_z$ subbands~\cite{Fang_1997_PRB}. The A$_{1g}$ CDW coherent oscillation of the folded Se-4p band also appears to have substructure that indicates the presence of multiple subbands~\cite{Sugawara_2016_ACSNano}. Fig.4(c) shows giant enhancement of fluctuations at x=0.04 where multiple electronic bands oscillate at the 6.2\,THz A$_{1g}$ coherent phonon frequency, as well as a dramatic smearing of the FFT at the 610\,nm probe resonance. At x=0.08 the disorder in the crystal suppresses much of the A$_{1g}$ amplitude except for the smeared phonon signal at the 610\,nm probe resonance.
On the other hand, the temperature dependence of the 640\,nm resonance corresponding to the Se-4p A band~\cite{Fang_1997_PRB} shows a decrease of the electron-phonon coupling close to the superconducting transition temperature T=4\,K (Fig.4(e)).
At constant fluence and assuming the electronic bands are no longer subject to renormalization, the high temperature optical responses (100\,K and 200\,K) exhibit a much larger electron-phonon coupling strength than those in the vicinity of the superconducting phase (3\,K and 6\,K) which contribute to the elevated $\tau_1$ decay constant in Fig.2.

Finally, we would like to note that one would expect the electron-phonon coupling to increase in strength near the superconducting phase transition, but this need only be true for those parts of the electronic band structure that contribute to the formation of Cooper pairs. In the superconducting Cu$_{0.08}$TiSe$_2$ it appears that the normal A$_{1g}$ LO phonon couples to the optically induced holes at the zone center (monitored by the 610\,nm resonance) rather than to the electrons that are dispersed at the zone edge (Fig.4(d,e)). So the electron-phonon coupling responsible for emergence of superconductivity is distinct from the one leading to commensurate CDW order.

In summary, the time resolved pump-probe optical spectroscopy results on Cu-doped TiSe$_2$ show that: (1) both the excitonic condensate and soft L$_1^-$ phonon contribute to formation of commensurate CDW order that is accompanied by PLD. Cu doping weakens the excitonic contribution to the CDW and decouples it from the L$_1^-$ phonon.  As a  result we have a lattice that is less susceptible to PLD through electronic excitation at x$\ge$0.04; (2) there is a strong enhancement of fluctuations in the folded Se-4p band in response to small optical perturbations near the quantum phase transition at x=0.04. At T$\sim$0 the system transitions from one with a strongly commensurate CDW phase at x$<$0.04 into the superconducting one at x$>$0.04 with coexisting weakened CDW; (3) there is a large electron-phonon coupling of the coherent A$_{1g}$ phonon to the Se-4p $\Gamma$ band for Cu$_{0.08}$TiSe$_2$ at low temperatures, accompanied by a reduced electron-phonon coupling to the Se-4p A band. In all other parts of the electronic band structure there is a loss of coherence in the A$_{1g}$ phonon signal with Cu intercalation due to the Cu-induced disorder.

Our results are in agreement with the earlier findings of Barath et al.~\cite{Barath_2008_PRL} in which Raman experiments on Cu$_x$TiSe$_2$ show onset of quantum fluctuations of the softened CDW-A$_{1g}$ phonon at x=0.04. The onset of fluctuations was taken as evidence for the transition to an incommensurate CDW phase at x$>$0.04. Indeed, previous STM measurements~\cite{Iavarone_2012_PRB} have shown that the CDW still coexists with the superconducting state up to at least Cu$_{0.06}$TiSe$_2$, meaning the transition at x=0.04 is not simply a transition from a CDW to superconducting phase. Instead, the enhanced fluctuations at x=0.04 and 3\,K presented in our work, show a decoupling of excitons from the soft L$_1^-$ phonon, and general loss of coherence and long-range order in the A$_{1g}$ phonon at higher copper doping. This points to a transition from commensurate to incommensurate CDW phase which is accompanied with the emergence of the superconducting state in a situation that is analogous to pure TiSe$_2$ under pressure~\cite{Snow_2003_PRL}. X-ray experiments show an incommensurate CDW in the vicinity of the superconducting phase~\cite{Joe_2014_NaturePhys,Kogar_arxiv2016}.

This material is based upon work supported by the National Science Foundation under Grant No. ECCS-1408151. The use of the Center for Nanoscale Materials, an Office of Science user facility, was supported by the U. S. Department of Energy, Office of Science, Office of Basic Energy Sciences, under Contract No. DE-AC02-06CH11357.

%
\end{document}